\pdfoutput=1
\documentclass[aps,prl,twocolumn,showpacs,superscriptaddress]{revtex4-1}
\usepackage{graphicx}
\usepackage{gensymb}
\usepackage{amsmath}

\begin{document}
\title{Edge Modes and Asymmetric Wave Transport in Topological Lattices: Experimental Characterization at Finite Frequencies}
\author{Jihong Ma}
\affiliation{Department of Civil, Environmental, and Geo- Engineering, University of Minnesota, Minneapolis, MN 55455, USA}
\author{Di Zhou}
\affiliation{Department of Physics, University of Michigan, Ann Arbor, MI 48109, USA}
\author{Kai Sun}
\affiliation{Department of Physics, University of Michigan, Ann Arbor, MI 48109, USA}
\author{Xiaoming Mao}
\affiliation{Department of Physics, University of Michigan, Ann Arbor, MI 48109, USA}
\author{Stefano Gonella}
\email{sgonella@umn.edu}
\affiliation{Department of Civil, Environmental, and Geo- Engineering, University of Minnesota, Minneapolis, MN 55455, USA}

\begin{abstract}
 Although topological mechanical metamaterials have been extensively studied from a theoretical perspective, their experimental characterization has been lagging. To address this shortcoming, we present a systematic laser-assisted experimental characterization of topological kagome lattices, aimed at elucidating their in-plane phononic and topological characteristics. We specifically explore the continuum elasticity limit, which is established when the ideal hinges that appear in the theoretical models are replaced by ligaments capable of supporting bending deformation, as observed for instance in realistic physical lattices. We reveal how the zero-energy floppy edge modes predicted for ideal configurations morph into finite-frequency phonon modes that localize at the edges. By probing the lattices with carefully designed excitation signals, we are able to extract and characterize all the features of a complex low-frequency acoustic regime in which bulk modes and topological edge modes overlap and entangle in response. The experiments provide unequivocal evidence of the existence of strong asymmetric wave transport regimes at finite frequencies.
\end{abstract}

\maketitle
Acousto-elastic metamaterials and phononic crystals are artificially architected materials endowed with the capability to manipulate mechanical waves. This attribute makes them attractive for a variety of emerging technological applications such as acoustic cloaking \cite{cummer2007one,chen2007acoustic}, sound manipulation and control \cite{cummer2016controlling}, smart sensing \cite{chen2014enhanced}, imaging \cite{craster2012acoustic}, and thermal management \cite{davis2014nanophononic}. An important dynamical property of periodic metamaterials is their inherent directionality, which manifests as a frequency-selective spatial anisotropy of their bulk wave modes. The directionality patterns are dictated by the symmetry of the unit cell. By relaxing the cell symmetry, e.g. through actively reconfigurable microstructural elements, it is possible to alter the global directivity of the medium \cite{celli2014low,Celli_Curlable_SMS}.

\begin{figure} [t]
	\includegraphics[scale=0.086]{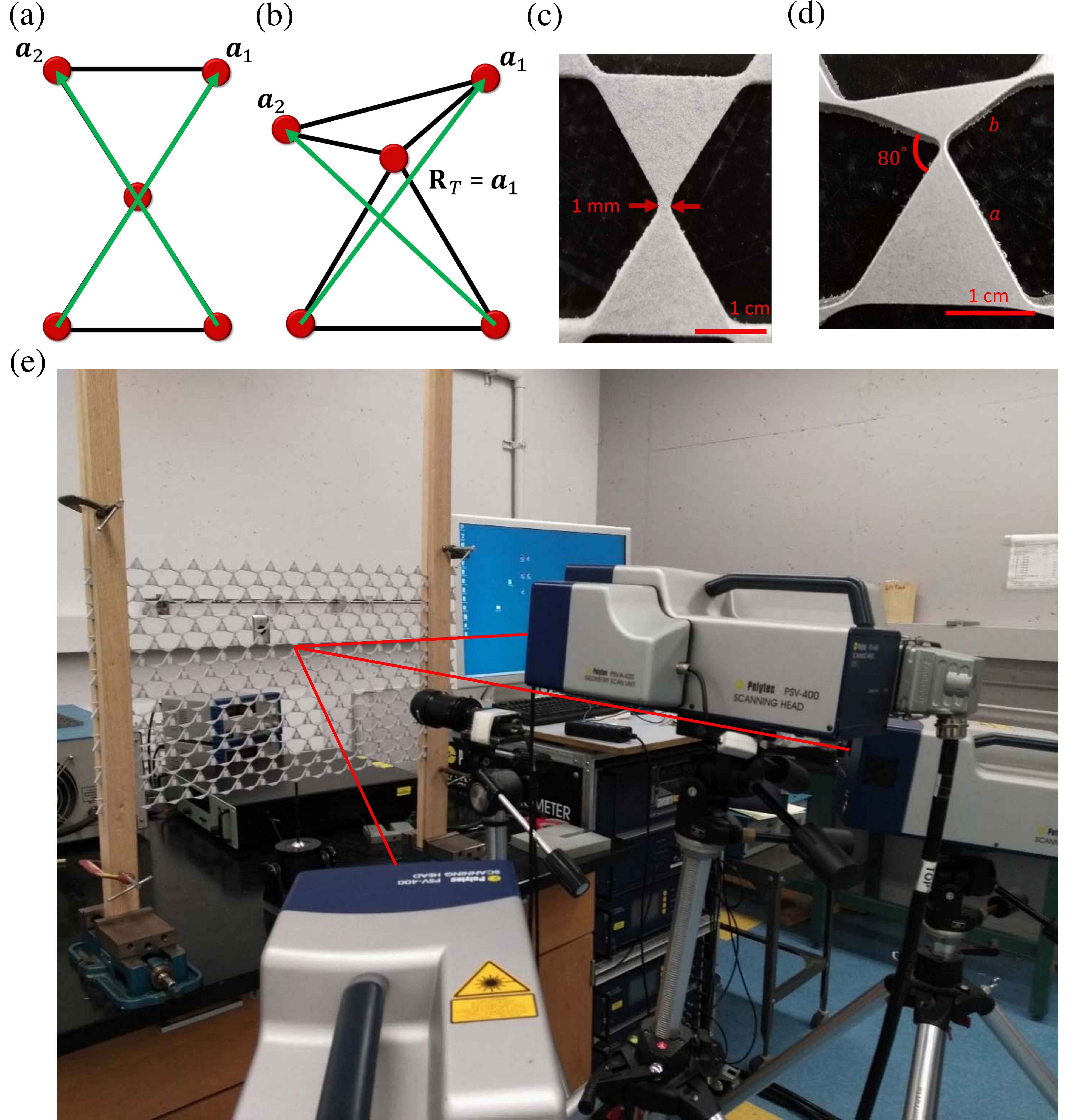} 
	\caption{(a)-(b): Unit cells of (a) regular kagome and (b) topological kagome lattices modeled as ideal ball-spring systems with lattice vectors $\mathbf{a}_1$ and $\mathbf{a}_2$ and polarization vector $\mathbf{R}_T$ indicated. (c)-(d): Unit cells of (c) regular kagome and (d) topological kagome lattices cut from ABS sheets and used in experiments. (e) Experimental setup showing a $8\times16$ topological lattice with the floppy edge at the top , excitation applied via shaker at the bottom edge and 3D laser vibrometer scanning heads.}
	\label{unitcell}
\end{figure}

Recent years have seen the advent of topological mechanical metamaterials, which have introduced a new paradigm for wave manipulation enabled by topologically protected wave modes. The concept originally emerged in the realm of quantum physics \cite{haldane1988model, kane2005quantum, hasan2010colloquium,qi2011topological}. For example, time-reversal invariant topological insulators are electronic materials that, in addition to their bulk bandgap behavior, possess topologically protected conducting edge states that are insensitive to defects \cite{hasan2010colloquium, qi2011topological}. Interestingly, these edge properties, which are only appreciated in finite domains, are controlled by topological invariants of the unit cell. Inspired by these intriguing properties, significant endeavors have been made to achieve mechanical analogues of these phenomena in the form of static \cite{kane2014topological, paulose2015topological, rocklin2017transformable, rocklin2016mechanical, stenull2016topological, bilal2017intrinsically} and dynamical \cite{susstrunk2015observation, nash2015topological, wang2015topological, mousavi2015topologically, kariyado2015manipulation, wang2015coriolis, pal2016helical, brendel2017pseudomagnetic, chaunsali2017demonstrating, chaunsali2018subwavelength, prodan2017dynamical} phonon edge modes. A special class of topological phenomena occurs in Maxwell lattices, i.e., frame structures featuring ideal hinges that allow free rotations of the struts \cite{maxwell1864calculation}. In Maxwell lattices, the number of constraints equals that of the degrees of freedom in the unit cell \cite{lubensky2015phonons}. The presence of free rotational mechanisms allows for lattice distortions that do not stretch the struts. This type of mechanism is referred to as a zero-frequency (energy) mode \cite{mao2018maxwell}. An example of a Maxwell lattice is the regular kagome lattice (unit cell shown in Fig.~\ref{unitcell}(a)). While an infinite lattice can be seen as perfectly constrained everywhere, a finite domain with a free boundary necessarily presents zero modes, whose number is proportional to the size of the boundary. Depending on the unit cell, these zero modes can manifest as plane-wave-like features in the bulk or localize at the boundaries as so-called floppy edge modes \cite{sun2012surface}.  

While the behavior of ideal topological lattices has been extensively studied theoretically using ball and spring models (and with a focus on the static response), experimental proofs of concept carried out on physical specimens (realized via fabrication techniques such as cutting, molding, or printing) have been rare \cite{paulose2015selective, bilal2017intrinsically}. In this letter, we attempt to bridge this gap in the literature through a suite of experiments conducted on physical topological specimens. The main source of complexity in going from ideal to physical lattices is that the singular hinge connections at the lattice nodes are replaced by finite-thickness ligaments capable of supporting bending deformation. We refer to these conditions as the continuum elasticity limit, to emphasize that the mechanical behavior of the hinges depends on the geometric properties of the ligaments and can be described by continuum elasticity models. 
Adding bending stiffness lifts the zero-frequency modes to finite frequencies, resulting in a complex acoustic regime where bulk and edge phonons are superimposed and entangled in the response. Our experiments essentially attempt to disentangle the coexisting wave components and interpret them in light of the geometric and topological descriptors of the lattice. In parallel, we visually reconstruct the spatial patterns of the edge modes that localize at the floppy boundaries and we demonstrate the existence of strong asymmetric wave transport regimes at finite frequencies. 

Reconstructing the phononic and topological characteristics of kagome lattices requires the capability to obtain in-plane measurements at points on the lattice surfaces. To this end, we use a 3D Scanning Laser Doppler Vibrometer (SLDV, Polytec PSV-400-3D), shown in Fig.~\ref{unitcell}(e). The lattice specimens are framed on the two sides, leaving the top and bottom edges free. The excitation is prescribed as a point force at the mid point of the bottom edge using an electrodynamic shaker (Bruel \& Kjaer Type $4809$, powered by a Bruel \& Kjaer Type $2718$ amplifier) placed at the bottom of the structure for in-plane excitation. The specimens are manufactured via water-jet cutting from a sheet of acrylonitrile butadiene styrene (ABS) with the following material parameters: Young's modulus $E=2.14 \, \textrm{GPa}$, Poisson's ratio $\nu=0.35$, density $\rho=1040 \, \textrm{kg}/\textrm{m}^3$. The unit cells of kagome lattice specimens corresponding to the configurations of Fig.~\ref{unitcell}(a) and (b) are shown in Fig.~\ref{unitcell}(c) and (d), respectively. The regular kagome unit cell is composed of two equilateral triangles connected with a ligament. The topological kagome unit cell contains one equilateral and one isosceles triangles. The side length of all the equilateral triangles for both lattices, and the longer side of the isosceles triangle in the topological lattice are all $a=2 \, \textrm{cm}$, while the shorter sides have length $b=2/\sqrt3\ \,\textrm{cm}$. The twist angle of the topological lattice is $10^{\circ}$ counter-clockwise, which makes the smaller angle between the triangles $80^{\circ}$. $\mathbf{a}_1$ and $\mathbf{a}_2$ are the lattice vectors in real space. The ligaments are approximately $2 \,\textrm{mm}$ in width. The specimens consist of  $8\times16$ arrays of unit cells. In order to excite the topological configuration from both the floppy and the non-floppy edges (which is necessary to probe the topological characteristics), we rotate the specimen by $180^{\circ}$ while keeping the excitation point fixed.

\begin{figure} [t]
	\includegraphics[scale=0.1]{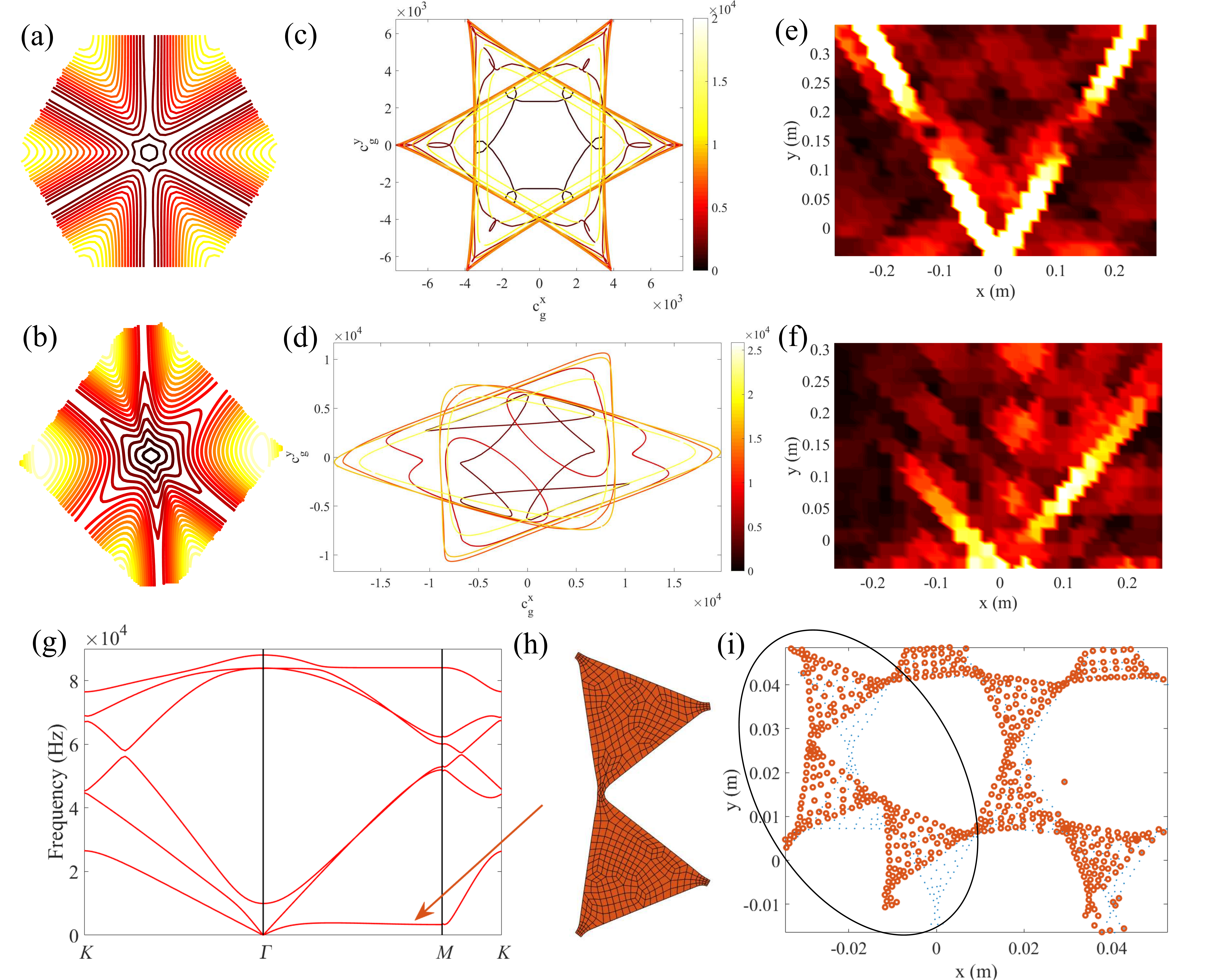} 
	\caption{(a)-(b): Iso-frequency contours (first phase-constant surface in the first Brillouin zone) of realistic (a) regular and (b) topological kagome lattices calculated from FEA. (c)-(d): Group velocity iso-frequency contours (frequency increasing from dark red to bright yellow) highlighting symmetric or asymmetric directivity of (c) regular and (d) topological configurations. (e)-(f): Snapshots of experimentally acquired wavefields in (e) regular and (f) topological lattices, confirming symmetric/asymmetric propagation. Dark red to bright yellow color denotes the  wave magnitude from small to large. (g) Band diagram for regular kagome lattice. (h) Mode shape calculated at the highlighted point at $\sim$ $3300 \, \textrm{Hz}$. (i) Detailed local scan focused on cells along the propagation path of the regular kagome lattice for excitation at $\sim$ $3300 \, \textrm{Hz}$. Dots and circles represent original and displaced positions of the scan points, respectively. The triangles inside the ellipse fall along the $\mathit{\Gamma-M}$ direction in real space.}
	\label{phonon}
\end{figure}

Our first preliminary task consists of experimentally reconstructing the bulk wave behavior. Beforehand we conduct a complete numerical characterization of the two unit cells of Fig.~\ref{unitcell}(c) and (d) using finite element analysis (FEA) and Bloch boundary conditions. The iso-frequency contours of the first phase-constant surface of the dispersion relation are presented in Fig.~\ref{phonon}(a) and (b). For completeness, the band diagram for the regular kagome structure is also plotted in Fig.~\ref{phonon}(g). Let us recall that band diagrams of ideal kagome lattices feature zero-energy modes involving displacements of lattice sites that occur without any deformation of the lattice \cite{mao2018maxwell}. Here, in contrast, as the physical hinges prevent the triangular plates from rotating freely with respect to one another, we do not observe any zero-energy mode (except for the two translational invariants), leaving the phonon spectrum fully gapped except at $\mathbf{q}=\mathbf{0}$, where  $\mathbf{q}$ is the wave vector in reciprocal space. Interestingly, the effect of finite-thickness hinges is analogous, in terms of added bending stiffness, to that obtained by adding next-nearest-neighbor interactions in ideal kagome lattices, discussed in Ref \cite{sun2012surface} and recalled in Fig.~\ref{supp_cite}. From a mechanics perspective, the rotation-hindering mechanisms are different between the two configurations: next-nearest-neighbor springs introduce additional elastic connections between selected nodes in the structure, while finite-width hinges only provide bending stiffness. However, both produce a shift toward finite frequencies of the isostatic zero-frequency branch in the $\mathit{\Gamma-M}$ direction and of the first optical branch at the $\mathit{\Gamma}$ point. To appreciate the frequency-selective spatial directivity intrinsic to these lattices, we track the evolution of the group velocity vector in the wave-vector plane by taking the gradient of the dispersion surface $\mathbf{\omega}$(${q_x, q_y}$) with respect to the Cartesian components of the wave vector, i.e., $\mathbf{c_g}=\nabla\mathbf{\omega}(q_x,q_y)$ and plotting iso-frequency contours of the gradient surface. Fig.~\ref{phonon}(c) and (d) show the group velocity contours for the first phonon mode of the regular and topological kagome lattices, respectively. As expected, the contour of the regular kagome lattice features a strict 6-fold symmetry, while the symmetry is broken in the topological case to reflect the intrinsic asymmetry of the unit cell. In essence, the effects of cell symmetry relaxation manifest in the spatial directivity landscape.

We proceed now to conduct experiments on the finite lattice specimens described above using the setup shown in Fig.~\ref{unitcell}(e). The laser scans performed for this task are fairly parsimonious, involving one measurement point per triangle. By exciting regular and topological kagome lattices with tone bursts with carrier frequencies falling in the acoustic bulk modes range, we recover the symmetric and asymmetric wave-fields shown in Fig.~\ref{phonon}(e) and (f), respectively. Both display wave beaming along the lattice vectors but markedly different degrees of symmetry, in conformance with the corresponding group velocity contours of Fig.~\ref{phonon}(c) and (d), respectively. The experimental data also matches the results of full-scale FEA simulations of finite lattices with identical size and characteristics, which are reported in Fig.~\ref{supp_FEA_full_scale}. By exciting the lattice with a broad-band chirp excitation, we can holistically reconstruct the entire phonon band diagram (in the frequency interval of the chirp), as shown in Fig.~\ref{supp_kagome_phonon}. Dedicated narrow band chirps spanning smaller frequency intervals can also be used to reconstruct partial sectors of the band diagram with higher fidelity. Finally, by performing local fine-scale scans, we can zoom in on selected details of the lattice cells to capture the localized deformation mechanisms of the hinges. For example, for a burst at $\sim$ $3300 \, \textrm{Hz}$ (which excites the plateau region of the first branch along the $\mathit{\Gamma-M}$ direction), in Fig.~\ref{phonon}(i) we plot the displacements of the scan points inside a few cells located along the corresponding direction in real space. Indeed, the deformation field presents a combination of transversal and rotational motion, consistent with the mode shape computed via unit cell analysis for the same conditions and shown in Fig.~\ref{phonon}(h).

While the bulk wave characteristics are fully captured via Bloch analysis, the topological edge modes, which manifest on the scale of finite lattices, require additional modeling tools. In the context of ideal models, the topological properties are often described by introducing the topological polarization vector ${\bf R}_T$, calculated as \cite{kane2014topological}:
\begin{eqnarray}
{\mathbf R}_T=-\sum_{i=1} n_i\mathbf{a}_i,
\label{polarization}
\end{eqnarray}
where $n_i$ are winding numbers obtained from 
\begin{eqnarray}
n_i=\frac{1}{2\pi i}\oint_{C_i}{d}\,\mathbf{q}\cdot\nabla_\mathbf{q}\,\mathrm{ln\,[det}(C(\mathbf{q})],
\end{eqnarray}
where $C_i$ is a cycle along the contour of the Brillouin zone connecting $\mathbf{q}$ and $\mathbf{q}+\mathbf{b}_i$, where $\mathbf {b}_i$ is a primitive reciprocal vector (such that $\mathbf {a}_i\cdot\mathbf{b}_j=2\pi\delta_{ij}$). $C(\mathbf{q})$ is the compatibility matrix of the ideal lattice describing the displacement-strain relation in Fourier space. The topological polarization vector reveals the existence of floppy edge modes, and identifies the specific edge(s) where floppy motion is exponentially localized. From Eq.~\eqref{polarization}, we can see that for the topological cell, ${\mathbf R}_T = {\mathbf a}_1$ and therefore points to the top boundary, thus qualifying it as a floppy edge. Although in this study the lattice connections are realistic ligaments, leading to restrictions on the possible in-plane bending between the plates, we conjecture that the existence of edge modes with floppy characteristics is preserved. The open question is how these modes would manifest in the dynamic response. In essence, we borrow the terminology ``floppy modes'' from the ideal lattice case and we broaden its significance to describe the more general localization of phonon modes that may occur at low (but finite) frequencies in their realistic counterparts. 

\begin{figure} [h!]
	\centering
	\includegraphics[scale=0.185]{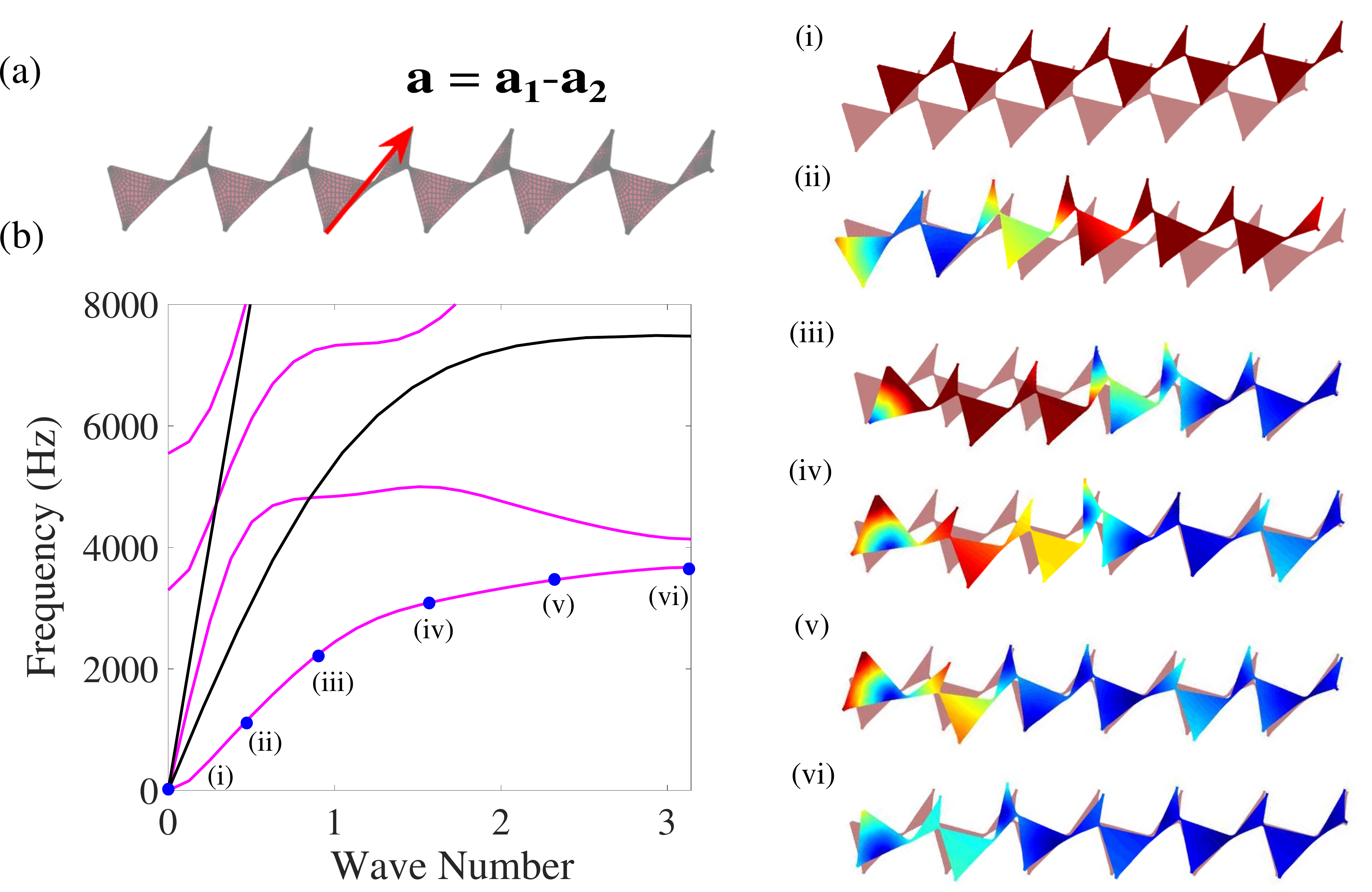} 
	\caption{(a) Supercell of topological kagome lattice comprising six unit cells. The left and right boundaries are free. Bloch conditions are applied along lattice vector $\mathbf{a}$. (b) Lowest branches of the band diagram for the supercell (magenta) superimposed to bulk modes calculated via unit cell analysis (black). (i)-(vi) Mode shapes of the supercell for different wave vectors sampled along the lowest floppy edge mode (blue circles). The color is commensurate to nodal displacement. The supercells are rotated by $45^{\circ}$ counter-clockwise comparing to the unit cell in Fig.~\ref{unitcell} (d).  }
	\label{supercell}
\end{figure}

Computationally, the fate of the floppy edge modes can be predicted through a \textit{supercell} analysis. To this end, we consider the finite strip of 6 cells shown in Fig.~\ref{supercell}(a), and we apply 1D Bloch conditions along vector $\mathbf{a}$, while leaving the top and bottom edges free to mimic an infinitely long topological strip. As can be seen from the band diagram calculated using FEA and shown in Fig.~\ref{supercell} (b), we recognize the emergence of two new branches, both confined in frequency intervals lower than the dispersive region of the lowest bulk modes. The mode shapes shown in Fig.~\ref{supercell}(i)-(vi), calculated along the lowest branch, feature indeed deformation localized at one end. This result qualifies these branches as edge modes and characterizes the left edge as the ``floppy'' boundary, consistent with the polarization vector predictions. These new modes represent the evolution of the floppy modes of ideal topological lattices in the limit of continuum elasticity. The fact that a topological lattice features two edge modes with distinct in-plane mechanisms constitutes an interesting departure from the conventional case of Rayleigh waves arising at the boundary of a 2D half-space, which feature a single mode of deformation \cite{rayleigh1885waves}.

%

\begin{figure} [t]
	\includegraphics[scale=0.2]{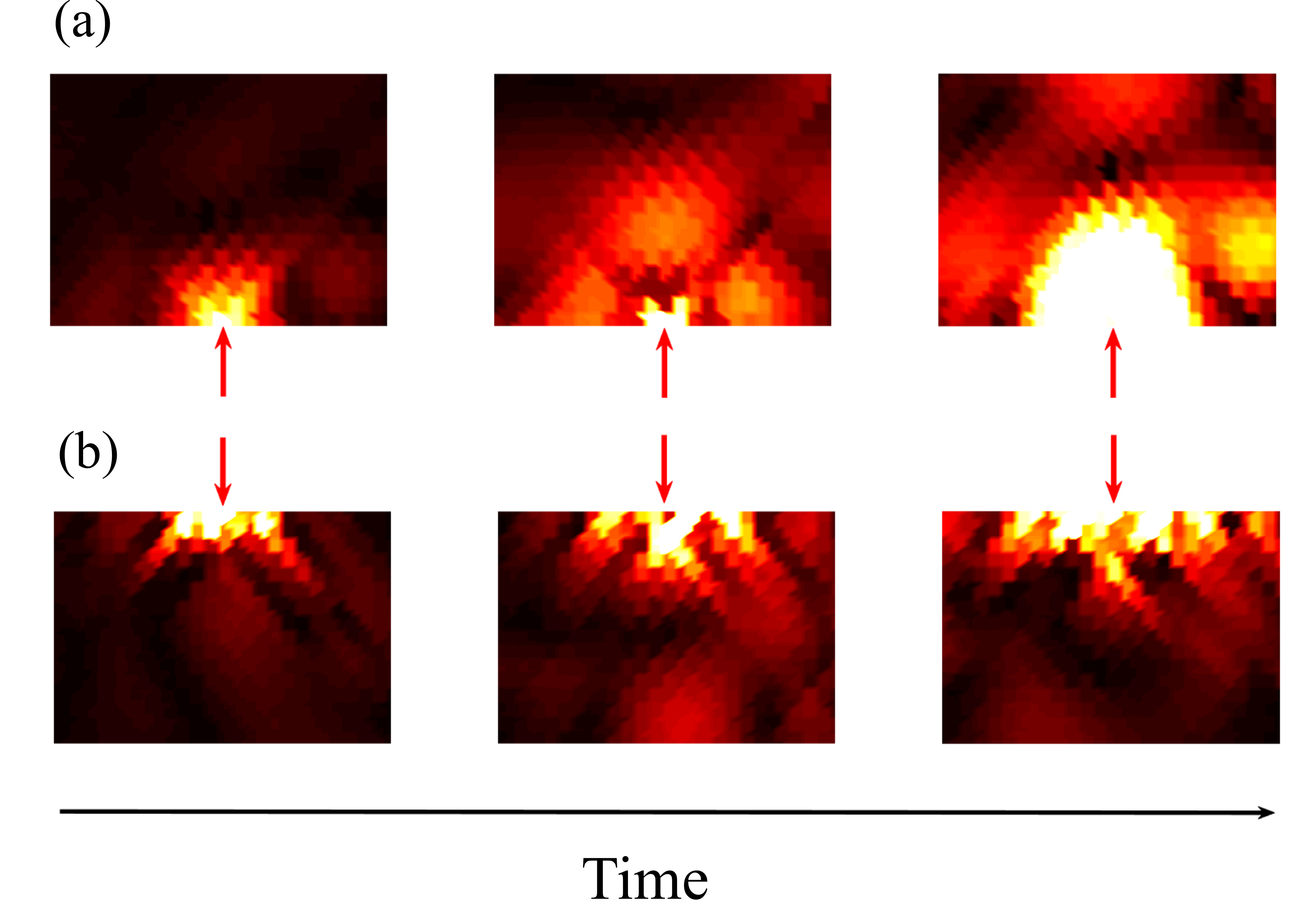} 
	\caption{Snapshots of wavefields induced through burst excitations at $\sim$ $3000 \, \textrm{Hz}$ applied at the (a) non-floppy and (b) floppy edges, respectively. In (b), waves mainly propagate along the floppy edge.}
	\label{propagation}
\end{figure}

\begin{figure} [b]
	\includegraphics[scale=0.156]{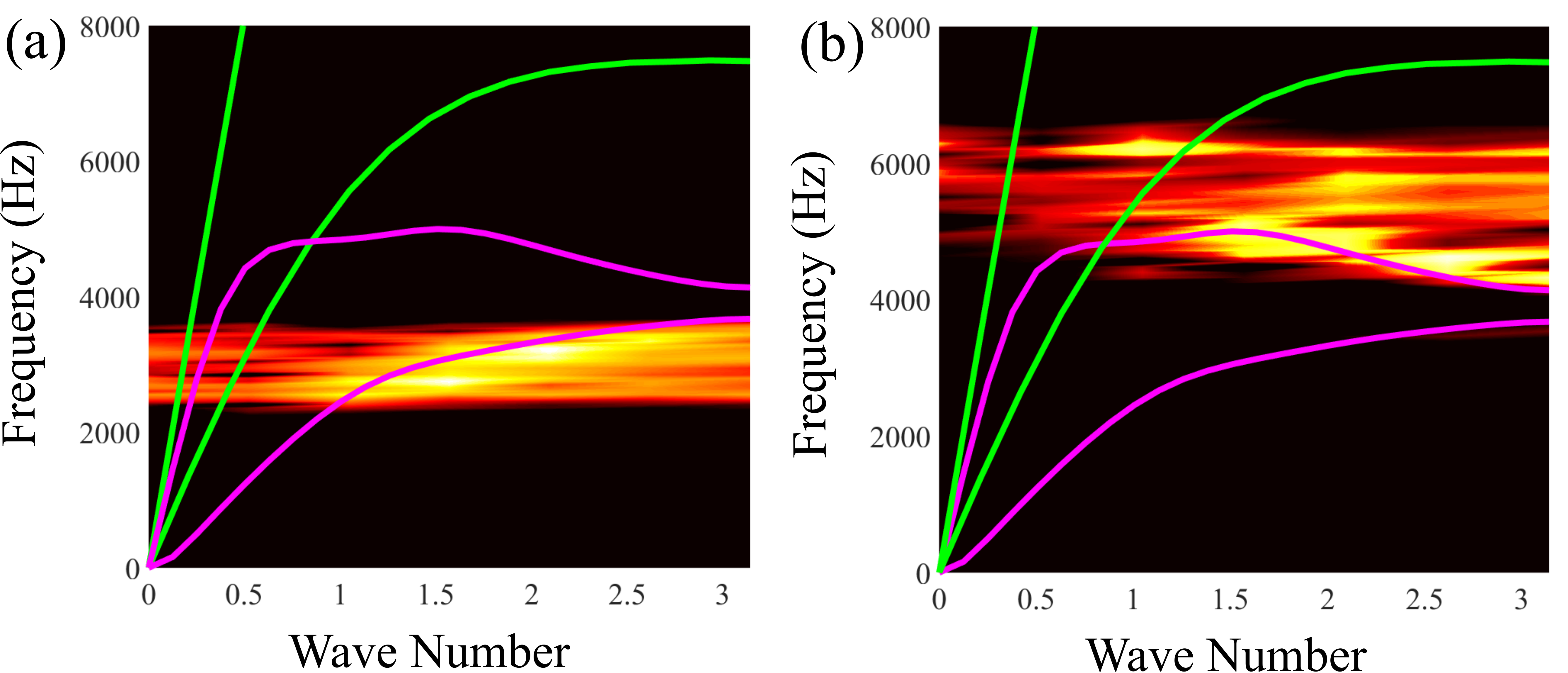} 
	\caption{DFT of experimental data for burst excitation at (a) $\sim$ $3000 \, \textrm{Hz}$ and (b) $\sim$ $5500 \, \textrm{Hz}$ applied at the floppy edge, matching the floppy branches from supercell analysis.} 
	\label{edgemode_band}
\end{figure}

To prove experimentally the existence of these localized modes, we resort to full scans of the specimen. When we excite at sufficiently low frequencies falling in the range of the edge modes (the two lowest magenta lines in Fig.~\ref{supercell}(b)), we capture the emergence of spatial behaviors that are germane to the topological configuration and that differ drastically according to which edge is excited (Fig.~\ref{propagation}). Specifically, by exciting from the non-floppy edge, we mostly observe waves that propagate fast in the bulk with very long wavelength. In contrast, by exciting the floppy edge, the energy remains localized at the boundary. These profound differences are reflected in the spectral plane upon discrete Fourier transform (DFT) of the data. Indeed, when we excite from the non-floppy edge, we detect the conventional signature of asymmetric phononic modes (reported for completeness in Fig.~\ref{supp_tkagome_phonon}). In contrast, when we excite from the floppy edge, most of the energy is concentrated in strong spectral features that align perfectly with the branches of the edge modes calculated via supercell analysis, as shown in Fig.~\ref{edgemode_band}.

\begin{figure} [h!]
	\includegraphics[scale=0.108]{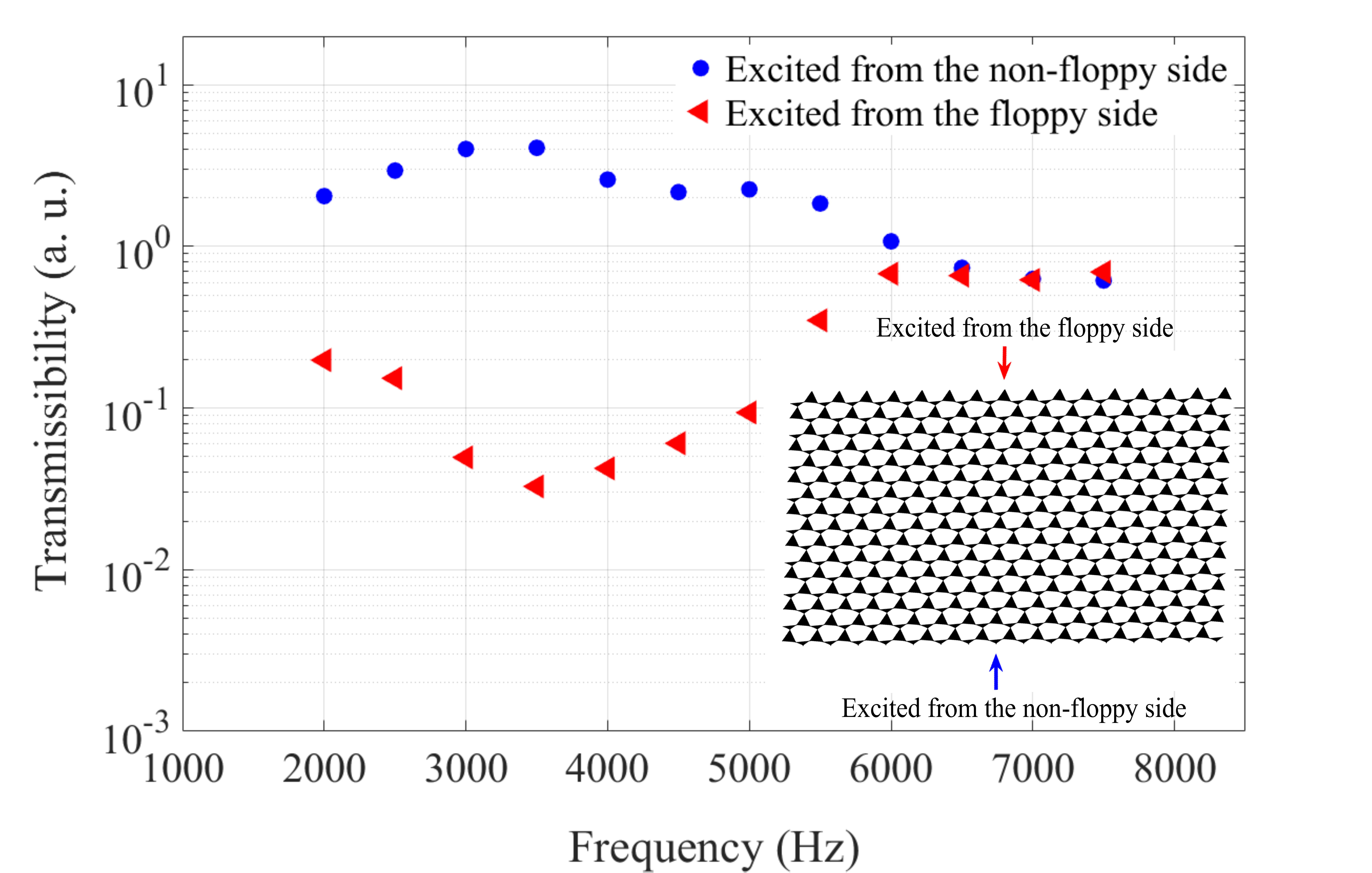} 
	\caption{Energy transmissibility as a function of frequency. The blue circles and red triangles correspond to excitations applied at the non-floppy and floppy edges, respectively.}
	\label{transmission}
\end{figure}

In an effort to quantify the asymmetry of wave transport that has emerged from these tests, we calculate the energy transmissibility, i.e., the ratio between the kinetic energy localized at the edge opposite to the excitation and the energy confined at the edge containing the excitation point. In Fig.~\ref{transmission}, we plot this quantity as a function of frequency for two scenarios: one where the excitation is applied at the non-floppy boundary, and the other for excitation applied at the floppy boundary. When we excite from the floppy edge at a frequency of $\sim$ $3300 \, \textrm{Hz}$, the transmissibility experiences a dip, which indicates that the energy is indeed localized at the floppy edge with modest propagation in the bulk. In contrast, for the same frequency, the excitation from the non-floppy edge results in a local maximum of the transmissibility, since the wave propagates through the bulk and reaches the opposite (floppy) edge, where it is eventually trapped by floppy mechanisms. However, when we excite above the edge mode frequency cut-off, the two transmissibility curves coalesce and we enter a conventional bulk wave regime.

In conclusion, we have experimentally demonstrated that by working in the limit of continuum elasticity, the phononic characteristics of topological kagome lattices are enriched by new edge modes at finite frequencies which can localize energy at the floppy boundary. This results in strong asymmetric wave transport capabilities in the very low-frequency acoustic regime. Our results help export a rich body of knowledge developed for static lattices to the dynamic regime, and provide innovative protocols for the dynamic experimental characterization of architected and lattice materials at large.

\vspace{0.2cm}
\section{acknowledgment}
	The authors acknowledge the support of the National Science Foundation (NSF grant EFRI-1741618).

%

\pagebreak
\widetext
\begin{center}
	\textbf{\large Supplemental Materials: Edge Modes and Asymmetric Wave Transport in Topological Lattices: Experimental Characterization at Finite Frequencies}
\end{center}
\setcounter{equation}{0}
\setcounter{figure}{0}
\setcounter{table}{0}
\setcounter{page}{1}
\makeatletter
\renewcommand{\theequation}{S\arabic{equation}}
\renewcommand{\thefigure}{S\arabic{figure}}

\section{\label{sec:level1}Phonon Band Diagram with Next-nearest-neighbor Interactions}

\begin{figure}[h!]
	\includegraphics[scale=0.18]{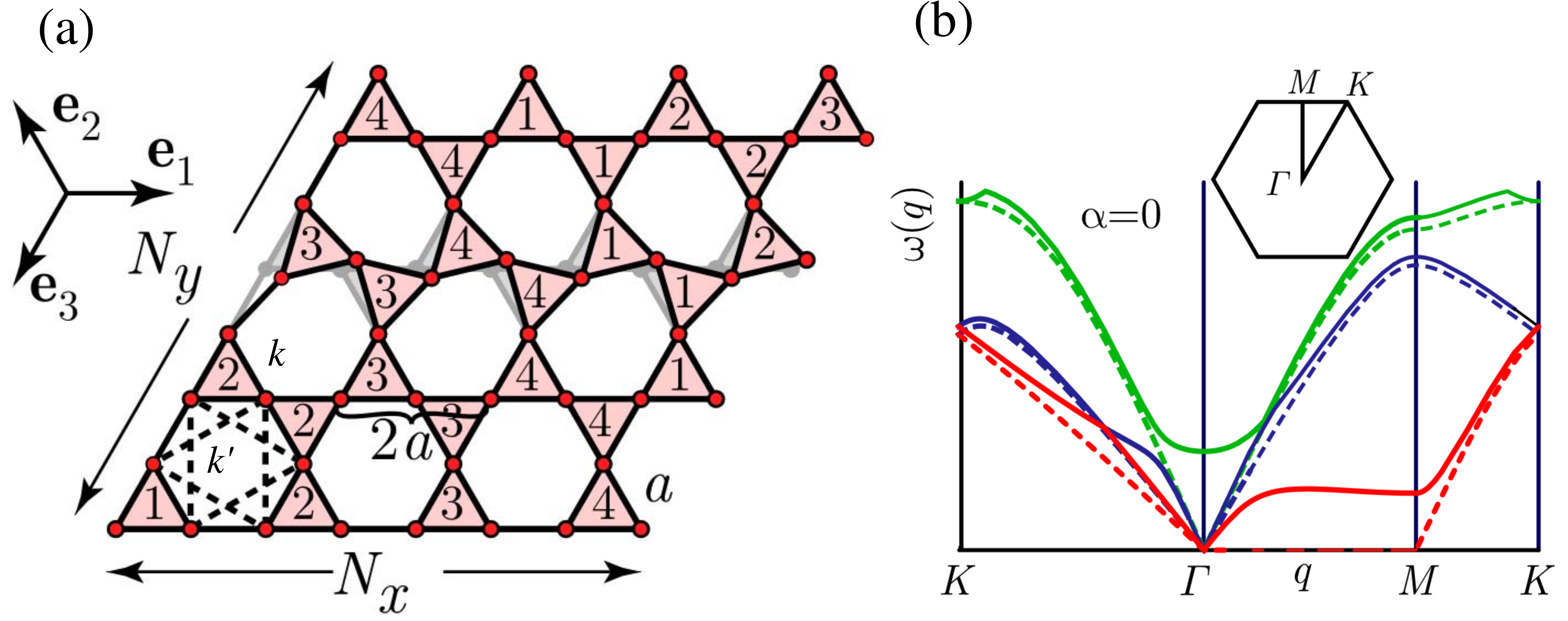} 
	\caption{(a) An ideal kagome lattice with nearest-neighbor springs with spring constant $k$ (black solid lines), and next-nearest-neighbor springs with spring constant $k'$ (dashed lines). (b) Band diagram. The dotted lines are for $k' = 0$ and the solid lines are for $k'= 0.02$. The isostatic and quasi-isostatic branches are shown in red. The figures are adapted from Ref.~\cite{sun2012surface}. }
	\label{supp_cite}
\end{figure}

In an ideal regular kagome lattice, when the interaction between lattice nodes is only through the main struts, i.e., when we only consider nearest-neighbor interactions with spring constant $k$, the triangles can freely rotate about their connection points, resulting in an isostatic zero-frequency branch in the $\mathit{\Gamma-M}$ direction. This is a zero-energy state, as the rotation of the plates does not result in any elongation or contraction of the struts.  In the presence of additional springs with spring constants $k'$ connecting the next-nearest neighbors (see Fig.~\ref{supp_cite} (a)), the zero-frequency branch is lifted to a finite-frequency range, as shown in Fig.~\ref{supp_cite} (b). The effect of $k' \neq 0$ is also felt on the first optical branch, where it has been shown that the non-zero optical frequency at the $\mathit{\Gamma}$ point is proportional to $\sqrt{k'}$,~\cite{sun2012surface}.

\section{\label{sec:level1}Full-scale finite element simulation}

\begin{figure}[h!]
	\includegraphics[scale=0.16]{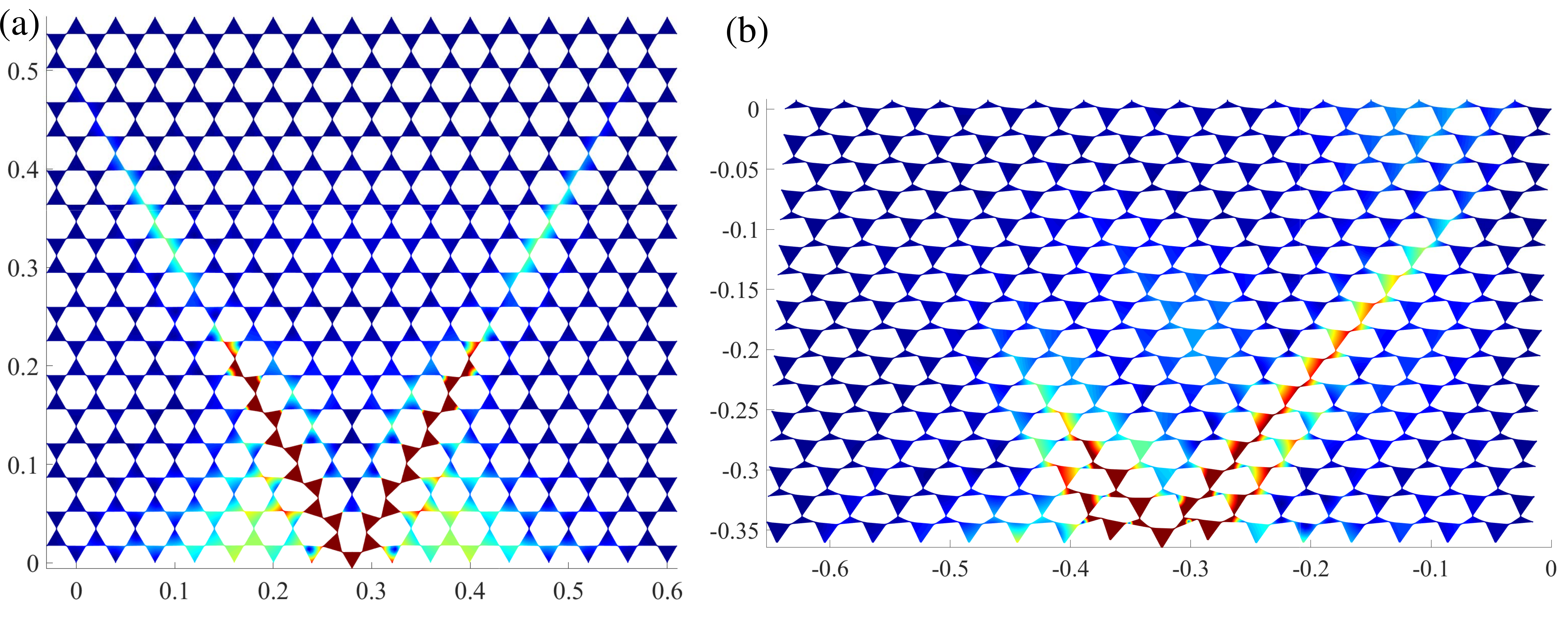} 
	\caption{Bulk wave propagation in (a) regular kagome lattice excited at  $\sim$ $8000 \, \textrm{Hz}$ and (b) topological kagome lattice excited at  $\sim$ $5500 \, \textrm{Hz}$, both calculated via full-scale finite element simulations.}
	\label{supp_FEA_full_scale}
\end{figure}

We conducted full-scale finite element simulations of a finite sheet of kagome lattice containing $8\times16$ cells. The left and right sides of the lattice are fixed. A burst point excitation with a carrier frequency at $\sim$ $8000 \, \textrm{Hz}$ is applied at the center of the bottom side of the lattice (see Fig.~\ref{supp_FEA_full_scale}(a)). The red-to-blue color map refers to nodal displacements (large to small). As we can see, the propagation of the wave presents a strong directionality, which conforms with the lattice vector directions. The simulation results are consistent with the experimental observations reported in Fig.~\ref{phonon}(e) in the letter. The simulation is repeated for the topological kagome lattice excited at $\sim$ $5500 \, \textrm{Hz}$. The results are shown in Fig.~\ref{supp_FEA_full_scale}(b) and are consistent with the experimental ones shown in Fig.~\ref{phonon}(f), revealing anisotropic wave beaming and symmetry relaxation.

\section{\label{sec:level1}Band diagram reconstruction via Fourier analysis}

The band diagram of a lattice can be reconstructed via Fourier analysis of spatio-temporal data acquired with the laser vibrometer. For this task, the excitation must be engineered carefully to probe the desired frequency intervals in which the branch reconstruction is desired. For the regular kagome lattice case, we collect time histories of velocities at measurement points located along the $\mathit{\Gamma-M}$ direction (the problem becomes one-dimensional in space with respect to this fixed direction), thus forming a matrix of velocity data. This matrix is then fed to a 2D discrete Fourier transform (DFT) with respect to time and space, which yields a matrix of spectral velocity amplitudes with respect to frequency and wave number. The DFT is calculated with the following expression:
\begin{eqnarray}
{V(k,f)=\sum_{n_1=0}^{N_1} \mathrm{e}^{-2\pi\mathit{i}k \, (n_1/N_1)}\sum_{n_2=0}^{N_2} \mathrm{e}^{-2\pi\mathit{i}f \, (n_2/N_2)} \, v(\chi,t)},
\label{dft}
\end{eqnarray}
where $v$ and $V$ are the amplitude of velocity along/perpendicular to the selected direction in real and Fourier space, respectively; $\chi$ is the position along the selected direction, $k$ is the wave number along such direction, $t$ is time, $f$ is frequency, and $N_1$ and $N_2$ are the total number of evaluation points in space and time, respectively. 2D color maps of $V(k,f)$ display the relation between frequency and wave number in the response. The dominant spectral features in these plots should conform to the branches of the band diagram calculated via unit cell analysis.

\begin{figure}[h!]
	\includegraphics[scale=0.19]{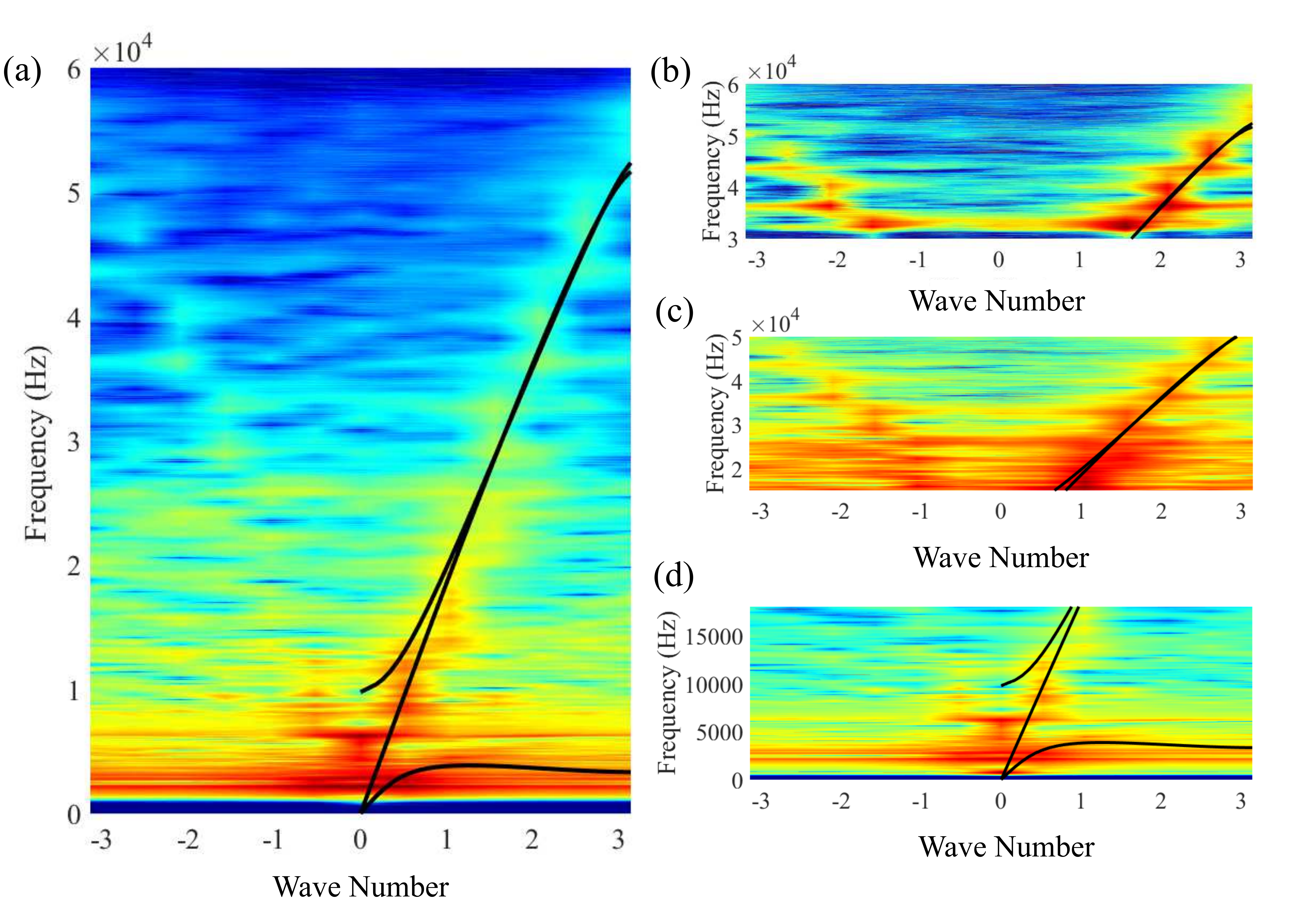} 
	\caption{Regular kagome band diagram reconstruction along  the $\mathit{\Gamma-M}$ direction obtained from experimental data via DFT. Comparison with branches calculated via finite element unit cell analysis (black curves). (a) Excitation with broad band chirp to globally reconstruct the first two bands. (b)-(d) Excitation with narrow band chirps for partial band reconstruction.}
	\label{supp_kagome_phonon}
\end{figure}

We excite the regular kagome lattice with a broadband chirp excitation with frequencies ranging from $100$ to $6\times10^4\, \textrm{Hz}$. As shown in Fig.~\ref{supp_kagome_phonon}(a), we can successfully reconstruct the first two phonon bands. 
To obtain more precise reconstruction and zoom in on the branch details, we use three narrow band chirps, and the reconstructed phonon bands are shown in Fig.~\ref{supp_kagome_phonon}(b)-(d). 

Similarly, we excite the topological kagome lattice from the \textit{non-floppy} edge with a burst excitation at $\sim$ $3000 \, \textrm{Hz}$ in order to compare the results with what observed for a similar excitation applied \textit{at the floppy edge} and reported in Fig.~\ref{edgemode_band}(a) in the letter. Here we compare the spectral features obtained via DFT with the branches of the band diagram obtained from unit cell analysis, which capture the bulk modes. We observe that overall, the transversal acoustic mode matches the computational branch more precisely than the longitudinal mode. 

\begin{figure}[h!]
	\includegraphics[scale=0.165]{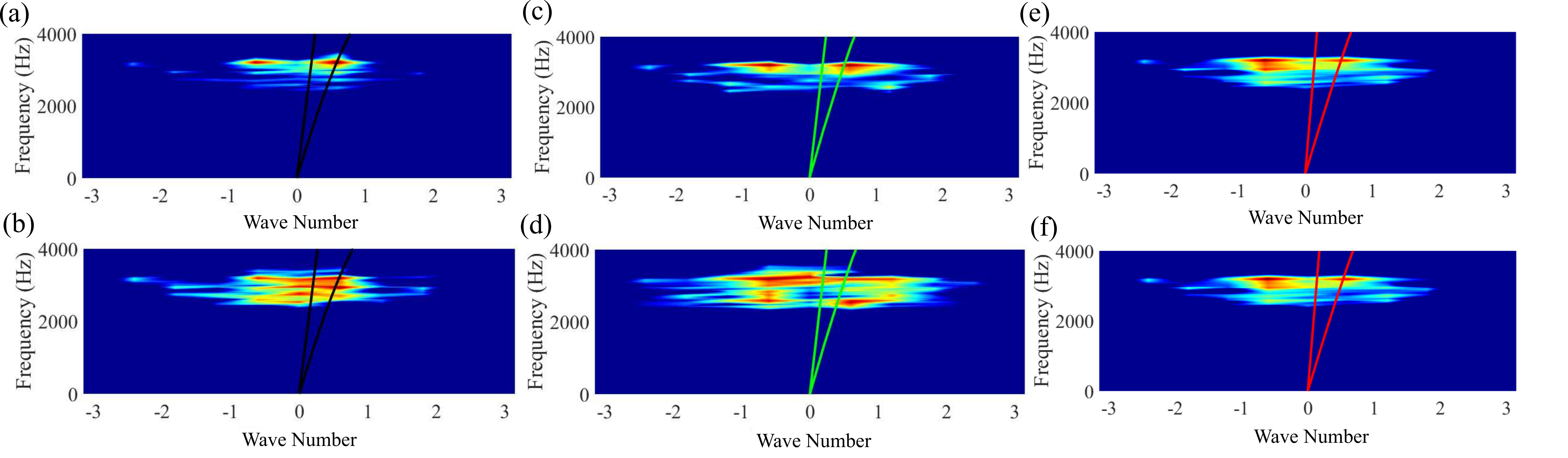} 
	\caption{Band diagram reconstruction for the topological kagome lattice from experimental data sampled along three lattice vectors. The excitation is a burst with carrier frequency of $\sim$ $3000 \, \textrm{Hz}$. (a), (c) and (e) show the reconstructed transversal acoustic modes along the $\mathbf{a}_1$, $\mathbf{a}_1-\mathbf{a}_2$, and $\mathbf{a}_2$ directions, respectively. (b), (d) and (f) show the reconstructed longitudinal acoustic modes along the $\mathbf{a}_1$, $\mathbf{a}_1-\mathbf{a}_2$, and $\mathbf{a}_2$ directions, respectively. The black, green and red curves represent phonon bands obtained from the unit cell analysis along the $\mathbf{a}_1$, $\mathbf{a}_1-\mathbf{a}_2$, and $\mathbf{a}_2$ directions, respectively.} 
	\label{supp_tkagome_phonon}
\end{figure}

\end{document}